# The New Horizons Kuiper Belt Extended Mission

**S.A. Stern**
Space Science and Engineering Division
Southwest Research Institute
1050 Walnut St., Suite 300
Boulder, CO 80302

**H.A. Weaver**
Johns Hopkins Applied Physics Laboratory
11100 Johns Hopkins Road
Laurel, MD 20723

**J.R. Spencer**
Space Science and Engineering Division
Southwest Research Institute
1050 Walnut St., Suite 300
Boulder, CO 80302

**H.A. Elliott**
Space Science and Engineering Division
Southwest Research Institute
6220 Culebra Road
San Antonio, TX 78238

**and the New Horizons Team**

**Corresponding Author:**
S. A. Stern
Space Science and Engineering Division
Southwest Research Institute
1050 Walnut St., Suite 300
Boulder, CO 80302
Ph: 303-324-5269, Email: astern@swri.edu

**Keywords**: NASA, New Horizons, Pluto, Kuiper Belt

# Abstract

The central objective of the New Horizons prime mission was to make the first exploration of Pluto and its system of moons. Following that, New Horizons has been approved for its first extended mission, which has the objectives of extensively studying the Kuiper Belt environment, observing numerous Kuiper Belt Objects (KBOs) and Centaurs in unique ways, and making the first close flyby of the KBO 486958 2014 MU$_{69}$. This review summarizes the objectives and plans for this approved mission extension, and briefly looks forward to potential objectives for subsequent extended missions by New Horizons.

# 1. Mission Background

New Horizons is a National Aeronautics and Space Administration (NASA) planetary exploration mission that resulted from a 2001 NASA competition for Principal Investigator-led Pluto-Kuiper Belt (PKB) exploration missions. NASA's call for proposals specified that the mission was to conduct a first reconnaissance flyby of Pluto, followed by subsequent flybys of one or more Kuiper Belt Objects (KBOs). In November 2001, New Horizons was selected as the winner of that competition. After the 2003 Planetary Decadal Survey (Belton et al. 2003) endorsed the funding of the PKB mission at high priority, and also endorsed a new class of PI-led missions in a considerably larger cost category than the Discovery program, New Horizons then became the first mission in NASA's then new ~$1B category planetary program of PI-led missions called New Frontiers.

The primary scientific objectives for New Horizons at the Pluto system were to map the surfaces of both Pluto and its large satellite Charon (at that time, no other satellites of Pluto were known), to map the surface compositions of Pluto and its large satellite Charon, and to determine the composition, pressure-temperature structure, and escape rate of Pluto's atmosphere. Secondary and tertiary objectives included higher-resolution geological and compositional studies of selected terrains, stereo mapping of terrain elevations, searches for trace species in Pluto's atmosphere, surface temperature measurements, the refinement of bulk parameters for Pluto and Charon, and searches for additional satellites and rings. Detailed information regarding these objectives can be found in Stern (2008) and Young et al. (2008). The instrument and spacecraft capabilities needed to complete these Pluto system objectives also enable New Horizons to collect key datasets about the Kuiper Belt and KBOs in extended missions.

The spacecraft is powered by a Radioisotope Thermoelectric Generator (RTG), which provides power at distances beyond where solar arrays are no longer effective. The spacecraft is capable of both spinning and 3-axis pointed operations. Data are stored on two 64-Gbit Solid State Recorders (SSRs) and are then transmitted to Earth via the spacecraft's X-band communications system. All attitude and trajectory maneuvers are carried out using a monopropellant hydrazine propulsion system that was launched from Earth with 78 kg of propellant; approximately 77 to 76 kg of this supply is useable (i.e., not trapped). More details regarding the spacecraft can be found in Fountain et al. (2009).

New Horizons carries a payload of 7 scientific instruments. These are: Ralph—a visible/IR remote sensing suite comprised of panchromatic and color imagers and an IR mapping spectrograph (Reuter et al. 2008); Alice—an ultraviolet mapping spectrograph (Stern et al. 2008); LORRI—a long focal-length panchromatic visible imager (Cheng et al. 2008); REX (Radio Experiment)—a radio science receiver (Tyler et al. 2008); SWAP (Solar Wind at Pluto)—a low-energy (~25-7500 eV) particle spectrometer (McComas et al. 2008); PEPSSI (Pluto Energetic Particle Spectrometer Science Investigation)—a high-energy (1-1000 keV) particle spectrometer (McNutt et al. 2008); and Venetia Burney SDC (Student Dust Counter)—a student-built team dust impact counter (Horanyi et al. 2008). Together this instrument suite provides these primary capabilities (Weaver et al. 2009): (i) medium- and high-resolution panchromatic visible wavelength mapping, (ii) medium-resolution visible wavelength color mapping, (iii) IR surface composition mapping, (iv) stereo imaging for terrain height mapping, (v) ultraviolet spectroscopy for atmospheric composition and vertical structure studies, (vi) plasma spectroscopy to measure the atmospheric escape rate and the composition of ionized gases escaping from Pluto's atmosphere and the interaction of the atmosphere with the solar wind, (vii) radio science to measure the brightness temperature of Pluto's surface, to determine the vertical temperature-pressure profile of Pluto's lower atmosphere, and to make bistatic radar measurements at 4.2 cm wavelengths, and (viii) a dust detector to search for particulates in orbit around Pluto. Additional details are given in Table 1.

## Table 1: New Horizons Instrument Payload Summary

| Instrument | Angular Resolution | Field of View | Wavelength or Energy Range | Notes |
|---|---|---|---|---|
| Ralph/MVIC | 19.8 µrad/pix | 99.3 mrad* | 400–550 nm<br>54 –700 nm<br>780–975 nm<br>860–910 nm<br>400–975 nm | Pushbroom color and panchromatic camera |
| Ralph/LEISA | 60.8 µrad/pix | 15.6x15.6 mrad | 1.25–2.5 µm | Near-IR imaging spectrometer: $\lambda/\Delta\lambda=240$ |
| LORRI | 4.96 µrad/pix | 5.08x5.08 mrad | 350 –850 nm | Framing camera |
| Alice | 1700 µrad/pix | 35x35 mrad;<br>1.7x70 mrad** | 47 –188 nm | UV imaging spectrometer: spectral resolution 0.18 nm |
| REX | 20 µrad | 20 mrad | 4.2 cm | Doppler tracking of uplink DSN signals; passive radiometry; interplanetary total electron column |
| SWAP | N/A | 276°x10° | 0.25 –7.5 keV*** | Low-energy plasma spectrometer |
| PEPSSI | 25°x12° | 160 x12° | 1–1000 keV*** | High-energy plasma spectrometer |
| SDC | N/A | 180 x180° | Mass>$10^{-12}$ g | Dust impact detector |

*Cross-scan FOV: along-scan FOV determined by scan length. A panchromatic framing mode, with 99.3 x 2.53 mrad FOV, is also available.** FOV consists of a "box" and a "slit," with the given dimensions *** Precise energy range depends on particle type.

New Horizons was launched on an Atlas V 551 rocket on 19 January 2006. It executed a Jupiter Gravity Assist on 28 February 2007 to target its Pluto flyby, which occurred at a heliocentric distance of 32.9 AU on 14 July 2015. The spacecraft's study of the Pluto system, including all five known moons and its interplanetary environment, spanned a six-month period from January through October of 2015. The New Horizons flyby of the Pluto system was fully successful (Stern et al. 2015), meeting and in many cases exceeding, the Pluto objectives set out for it by NASA (Lunine et al. 1997) and the National Academy of Sciences (Belton et al. 2003).

Following the success of the Pluto flyby, and with a fully operational and healthy spacecraft and payload, the New Horizons project proposed a 5-year long Kuiper Belt Extended Mission (KEM), which was reviewed and then formally accepted on 1 July 2016. KEM extends spacecraft exploration of the Kuiper Belt out to a distance of 50 AU, which the spacecraft will reach in early 2021. This review details the objectives and plans for that 2016-2021 extended mission, and then briefly considers the potential for future extended missions for New Horizons.

# 2. KEM Overview and Objectives

The Kuiper Belt (KB) is a rich scientific and intellectual frontier. Its exploration has important implications for better understanding comets, the origin of small planets, the solar system as a whole, the solar nebula, and extra-solar disks, as well as for studying thermally primitive material from the planet formation era. This exploration by New Horizons will transform KB and KBO science from a purely astronomical regime, as it was previously, to a geological and geophysical regime. This exploration has been strongly endorsed by two decadal surveys (Belton et al. 2003; Squyres et al. 2011), as well as the Pluto-KB Mission Science Definition Team (Lunine et al. 1997).

The centerpiece (and threshold mission) of KEM is the close flyby of a "cold classical" KBO called (486958 2014 $MU_{69}$, hereafter MU69); this flyby will reach closest approach early (UTC) on 1 Jan 2019. The KEM flyby of MU69 will obtain the first and only planned high-resolution geological and compositional studies of a small KBO and will make the first sensitive searches for coma activity and satellites/rings around any small KBO.

MU69 was discovered using the Hubble Space Telescope (HST). MU69's orbit is now well determined from intensive HST astrometry over a 3.5-year arc, yielding a semi-major axis, eccentricity, and inclination near 44.5 AU, 0.04, and 2.5 degrees, respectively. Little else is known about MU69 except (i) its color (0.36±0.38 magnitudes redder than the Sun and ~0.3 magnitudes redder than Pluto; Benecchi et al. 2018); (ii) its lightcurve amplitude (constrained to be <20%, indicating either a crudely equi-dimensional shape or a pole vector approximately aligned to its current line of sight to Earth, or both; Benecchi et al. 2018); and (iii) some information learned about its size, shape, and albedo from three stellar occultations in 2017 (Buie et al. 2018).

This ~30 km diameter KBO (absolute magnitude ~10.9), which may be a binary or contact binary (Buie et al. 2018) is ~$10^3$ times more massive than comet 67P/Churyumov-Gerasimenko, which the ESA/NASA Rosetta mission orbited, but is ~$5\times10^5$ times less massive than Pluto. This mass places MU69 in a key intermediate size regime to increase our understanding of planetary accretion. Its orbit places it firmly in the Cold Classical KB. Given this information, MU69 is believed to have existed near its 44 AU semi-major axis for 4+ Gyr, in storage at ~35 K. As such, MU69 will be the most primitive body ever studied by any spacecraft.

New Horizons (NH) made propulsive maneuvers in late 2015 to target the MU69 flyby. The planned flyby will approach MU69 to ~3,500 km, ~3.5 times closer than NH flew past Pluto; consequently, imaging and compositional mapping spectroscopy resolutions will be ~2x enhanced over what was obtained at Pluto. The reason that the resolution is not improved proportionately with the closer approach distance has to do with when in the flyby timeline the various measurements have been scheduled.

In this flyby, ~50 Gbits of MU69 data will be collected (cf. ~55 Gbits at Pluto). Owing to MU69's greater range to Earth than Pluto, the downlink of the MU69 data will take ~21 months, i.e., through approximately September of 2020. Data analysis, publication of the initial results, and Planetary Data System (PDS) archiving is planned to be complete by September 2021.

The New Horizons KEM, however, is much richer than just the close flyby of MU69. As described below, it also comprehensively exploits the unique resource of New Horizons as an observation platform in the KB. Additionally, it is worth pointing out that New Horizons is the first spacecraft intentionally sent to study the Kuiper Belt and objects in it, as both Pioneers and both Voyagers entered this region before the Kuiper Belt was discovered.

The remainder of this paper is organized as follows: In §3 below, we describe sensitive, and nearly continuous, post-Voyager measurements of the KB/heliospheric dust and plasma environment across the KB, that KEM will achieve. In §4 below, we describe the various unique studies of KBOs and Centaurs (escaped KBOs) that New Horizons is conducting in KEM. In §5 below, we then describe the MU69 flyby in detail, including its scientific objectives and observing plan. In §6 below, we go on to discuss the potential for important additional science New Horizons can conduct in subsequent extended missions. §7 provides a brief summary.

## 3. New Horizons as an Observatory in the Kuiper Belt: The Kuiper Belt Heliospheric Transect

The primary goal of the KEM heliospheric observations is to conduct a transect of the Kuiper Belt from 33 to 50 AU to characterize the dust, neutral gas, solar wind, and energetic particle environments there. The 50 AU distance was chosen to correspond to the aphelion of Pluto. As a result, in addition to understanding the Kuiper Belt's charged particle, neutral gas and dust environments for heliospheric science purposes, these heliospheric observations will also inform how space weathering affects the surfaces of Pluto and KBOs. These observations advance objectives to understand the outer heliosphere, and its interaction with the very local interstellar medium (VLISM) by measuring populations not previously measured, and by providing important new statistics under different solar cycle conditions. These observations are also being collected both while solar wind observations are being measured at 1 AU and images of the outer boundaries of the heliosphere are being taken with IBEX, something not possible when the Voyager and Pioneer spacecraft sampled these heliocentric distances.

Prior to New Horizons, only the Pioneer 10 and 11 and Voyager 1 and 2 missions had explored the outer heliosphere beyond 30 AU (see Figure 1). Although the Pioneers and Voyagers have gone beyond 30 AU, their solar wind plasma observations are more sparse than the mission trajectory plots would suggest because the Pioneer 10 and 11 observations only extend to 63.0 and 35.6 AU, respectively, and the solar wind plasma instrument on Voyager 1 stopped working near Saturn at ~9.74 AU. Voyager 2 has obtained the most extensive observations of the solar wind in the outer heliosphere extending to ~83 AU when Voyager 2 began to observe the foreshock of the termination shock and then crossed the termination shock at ~84 AU (Richardson & Stone 2009). With so few previous observations the NH observations help us understand both what is typical in the outer heliosphere and the long- term variations that depend on the solar cycle activity level.

Further, because of the measurement capability limitations of early mission instruments, some key charged particle populations have been inaccessible to direct observation in this heliocentric range and beyond, primarily due to energy spectrum gaps and sensitivity limitations (e.g., Belcher et al. 1980). The SWAP (McComas et al. 2008) keV particle spectrometer and PEPSSI (McNutt et al. 2008) MeV particle spectrometers aboard New Horizons provide crucial, missing energy coverage from ~6 to 20 keV with much higher sensitivities than the analogous Voyager plasma spectrometer (see Bridge et al. 1997) and Voyager's Low-Energy Charged Particle (LECP; see Krimigis et al. 1977). Therefore the New Horizons observations by PEPSSI and SWAP extend the charged-particle heliospheric observations of previous missions by measuring all populations simultaneously. This enables determining how much the suprathermal populations (e.g. interstellar pickup ions & suprathermal energy tails) modify the interplanetary shocks and the outer boundaries of our solar system. The NH and V2 observations are consistent with one another and most of the differences seem to be owing to differences in the solar activity level.

The Voyager 2 observations indicate that the plasma temperature rises beyond 30 AU (Richardson & Smith 2003) and that by 60 AU the solar wind speed is reduced by ~13% compared to its speed at 1 AU (Richardson & Wang 2003). Elliott et al., (2016) concluded that the average solar wind speed at New Horizons was already starting to decrease owing to the interaction with the interstellar material after examining the New Horizons solar wind observations from 11- 33 AU). Even more recent observations indicate this speed decrease with increasing distances is clear beyond 33 AU.

An oscillation in the speed with an amplitude of ~30 km/s and a period of ~2 days developed at ~48 AU in the Voyager 2 data (Paulaarena et al 1996; Zank et al. 1996)) simulated these oscillations as resulting from charge exchange in regions of high and low speed streams producing a pressure imbalance when thermal protons are converted to suprathermal pickup ions. SWAP has the capability to measure these oscillations. Elliott et al. (2016) has already used SWAP observations to show that power in the solar wind parameters moves from shorter to longer periods as solar wind features are worn down and merge with increasing distance from the Sun (Elliott et al. 2016). Understanding the overall pressure and energy flux budget for the solar wind, pickup ions and suprathermal ions is critical for modeling the outer heliosphere and for inputs to models simulating the IBEX images. McComas et al. (2017) has done this pressure analysis of SWAP observations out to 38 AU. Coincidently, New Horizons is headed along the same longitude as the Voyager 2 mission, but is in the ecliptic and Voyager 2 went well South of the reaching a latitude of -27.5° when it crossed the termination shock. Voyager 2 went South of the Energetic Neutral Atom (ENA) "ribbon" observed by IBEX (McComas et al. 2009). New Horizons is headed towards the ribbon and observing solar wind headed towards the ribbon and the NH Solar Wind Around Pluto (SWAP) solar wind observations provide a new critical outer heliospheric data set needed to test "ribbon" models.

Little is known about the dust environment and neutral H distibution in the outer heliosphere. The NH Student Dust Counter (SDC) is the first dedicated and calibrated dust instrument to measure the density and size distributions of interplanetary dust particles (IDPs) beyond 18 AU (Horanyi et al., 2008. This is because the dust detectors on both Pioneers failed by 18 AU and no systematic neutral H measurements were made far from the Sun by either the Pioneer or Voyagers. In contrast, dust impact fluxes are being measured by the Student Dust Counter (SDC) aboard New Horizons (Horanyi et al. 2008) to 50 AU in KEM. Neutral hydrogen measurements are also being made using the New Horizons Alice UV spectrograph (Stern et al. 2008) to 50 AU; and the total electron content (i.e., integral column from NH to Earth) are being measured to 50 AU using the New Horizons REX instrument (Tyler et al. 2008).

Regarding the heliosphere's neutral gas, the Alice instrument is being used once or twice annually (i.e., every ~1.5-3.0 AU in distance) to map the distribution of neutral hydrogen, the dominant neutral gas in the distant heliosphere. These observations provide critical inputs to understand and simulate the outer boundaries of the heliosphere. The full-slit Alice sensitivity at the 1216 Å wavelength of the hydrogen Lyman alpha (Lyα) line is ~550 counts/s/Rayleigh, so even a 1-Rayleigh (R) signal can be measured to 10σ accuracy in only 0.2 s. Figure 2 shows a set of sample Lyα measurements made by Alice compared to models (Gladstone et al. 2017). These observations extend those of Voyager (Quémerais et al. 2010), and provide valuable information for understanding how the solar wind and interstellar medium interact, as well as for understanding IBEX mission global Energetic Neutral Atom (ENA) maps (McComas et al. 2017).

Regarding heliospheric dust measurements, SDC is the first dedicated dust instrument to measure the spatial density and particle size distributions of interplanetary dust particles (IDPs) beyond 18 AU (Horanyi et al. 2008; see also Figure 3 here). These measurements improve on Voyager's complementary, but indirect observations of dust densities derived from its Plasma Wave System (PWS) (e.g., Gurnett et al. 1997, 2005). It has been suggested that the KB may be a primary source of Interplanetary Dust Particles (IDPs) (Landgraf et al. 2002) produced by collisions between KBOs (e.g. Stern 1996) and the bombardment of KBOs by IDPs and Interstellar Dust Particles (IDPs; e.g., Yamamoto & Mukai 1998). Recent models of KB dust production and loss use Pioneer 10 and 11 and NH observations to estimate the strengths of the dust sources and sinks, and follow dust transport in the outer Solar System (Vitense et al. 2014; Poppe 2016). However, the two predicted dust fluxes (again, see Figure 3) dramatically diverge beyond ~35 AU (see Poppe (2016) and Piquette et al. (2017)) because Poppe (2016) included an additional population of "outer/detached" KBOs. Therefore, by directly measuring the Kuiper Belt dust environment for the first time, SDC can discriminate between these divergent model results, and will also determine the dust production rate of the KB and its variation with heliocentric distance; these data also enable better comparisons of Kuiper Belt dust populations to dust disks around other stars.

The specific objectives of these heliospheric measurements, and the New Horizons instruments which address each are defined as follows:

1) Determine how much the solar wind is slowed and heated owing to interaction with interstellar material (SWAP and PEPSSI).

2) Measure the radial evolution of the suprathermal ion and energetic particle energies (PEPSSI).

3) Explore the evolution of the solar wind in the outer heliosphere, development of Merged Interaction Regions (MIRs) (Witasse et al. 2017; Elliott et al. 2016), and the modification of shocks by suprathermal ions (SWAP).

4) Characterize the response of pickup ions, energetic particles suprathermal tails to shocks and MIRs in the outer heliosphere (PEPSSI and SWAP).

5) Determine how the periodicities in the solar wind evolve with distance (SWAP).

6) Measure the particle pressures and energy fluxes of the solar wind, pickup ions and suprathermal tails in the Kuiper Belt region of the heliosphere (SWAP and PEPSSI).

7) Measure the distribution of cosmic rays (particularly the galactic cosmic rays) in the outer heliosphere (PEPSSI).

8) Determine the distribution of electrons across the Kuiper Belt (REX).

9) Measure the Ly-α distribution in the Kuiper Belt to map the interstellar H neutrals, which interact with heliosphere (Alice).

10) Determine the distribution of dust across the Kuiper Belt (SDC).

11) Address whether there is an extended source of dust in the Kuiper Belt (SDC).

Among the other notable aspects of New Horizons' heliospheric studies are:

- PEPSSI provides much needed measurements on the variation of a broad range of particle populations: the He pickup ions (>~2 keV/nuc), energetic particle H, He, and O (30 keV to 1 MeV), and cosmic ray H (>85 MeV, with coarse energy resolution up to ~1.5 GeV) populations in the outer heliosphere. PEPSSI's coincidence detection method eliminates backgrounds, allowing it to measure pickup $He^+$ and suprathermal particles at fluxes and energies between 2-20 keV, which are inaccessible to Voyager instruments. Although some of these populations measured by PEPSSI were also measured by Voyager (Decker & Krimigis 2003), Voyager made no composition measurements below ~1 MeV. PEPSSI is providing the first compositional (H, He, O) observations of the charged particle populations in the Kuiper Belt region using its detection capabilities from 30 keV to 1 MeV. PEPSSI also provides cosmic ray (primarily the Galactic Cosmic Rays (GCRs)) observations for H from 85 to MeV to 1.5 Gev, which determine how GCRs in the outer heliosphere vary with the solar cycle and heliospheric distance. PEPSSI is now the only cosmic ray monitor in the outer heliosphere because Voyager 1 and 2 are either in the interstellar medium or deep in the heliosheath (e.g., Richardson et al. 2017; see also https://voyager.jpl.nasa.gov/mission/status/).

- New Horizons is traveling along the same heliospheric longitude as Voyager 2, however New Horizons is in the ecliptic, whereas Voyager 2 is travelling south of the Energetic Neutral Atom (ENA) "ribbon" observed by IBEX (McComas et al. 2009). New Horizons is headed towards this ribbon and is observing solar wind traveling towards this ribbon.

- Both the SWAP and PEPSSI instruments observe interplanetary shocks, as shown in Figure 4. In particular, SWAP observes MIRs in the bulk solar wind plasma, associated shocks (Elliott et al. 2016), corresponding changes in the interstellar pick-up $H^+$, and even occasional interstellar pickup $He^+$ (McComas et al. 2017; Randol et al. 2012; 2013).

- PEPSSI observes interstellar pickup of $He^+$ and heavier species (Bagenal et al., 2016). PEPSSI observations also enable the study of how the suprathermal tails, $He^+$ and $O^+$ interstellar pickup ions, energetic particle, and GCRs respond to Globally Merged Interaction Regions (GMIRs).

- REX provides solar wind electron density fluctuations integrated over line-of-sight path between Earth and NH by examining intensity changes in the uplink radio signals, similar to interplanetary scintillation. REX is operated approximately monthly for this purpose.

Early results of the KEM heliospheric studies have been reported by Bagenal et al. (2016), Elliott et al. (2016), Gladstone et al. (2016), McComas et al. (2016; 2017), and Piquette et al., (2017). Regarding these results, Figure 5 shows an example distribution of distinct $H^+$ and $He^{++}$ solar wind peaks, a clear $H^+$ interstellar pickup ion cutoff, and a portion of the $He^+$ interstellar pickup distribution (McComas et al. 2017). Elliott et al. (2016) concluded that the average solar wind speed at New Horizons has started to decrease owing to interaction with interstellar material.

## 4. New Horizons as an Observatory in the Kuiper Belt: Context Surveys of Other KBOs and Centaurs

A second aspect of the KEM mission "observatory" science is that it takes advantage of the spacecraft's unique location *within* the KB to enable the LORRI telescope/imager investigation of 24-35 Distant KBOs (DKBOs) and Centaurs from uniquely high solar phase angles unattainable from Earth or any other operating spacecraft imaging system, and with spatial resolutions at those bodies often exceeding what is achievable from any other observatory, including both the HST and the James Webb Space Telescope (JWST). Careful calculations demonstrated that, unfortunately, no DKBOs can be detected with the other remote sensing instruments aboard New Horizons, so no UV, IR, thermal brightness, or visible color measurements are feasible.

**Distant KBO Context Surveys.** LORRI will make four different types of DKBO survey measurements: (1) light curves at multiple aspect (i.e., viewing) angles to determine shapes, rotation rates, and pole positions that cannot practically be obtained from Earth; (2) photometry at multiple phase angles to determine phase functions and photometric properties/regolith microphysical properties that cannot be obtained from Earth or elsewhere in the inner solar system; (3) sensitive deep-imaging searches for both smaller satellites and closer binaries that cannot be achieved using telescopic groundbased adaptive optics systems, HST, or JWST; (4) high phase photometric searches for ring and dust material around KBOs and Centaurs; and (5) astrometry measurements of each observed DKBO to improve orbital solutions for these bodies so that they can be more confidently studied with future capabilities expected to result from extremely large groundbased telescopes becoming operational in the 2020s.

From these observations, New Horizons KEM will yield a statistical sample of KBO surface properties, shapes, and binarity that is only possible from investigations taken within the KB. Because KEM will study both (1) nearby small KBOs (similar in size class to 2014 MU69) and (2) distant dwarf planets, KEM will better place MU69, Pluto, and Charon in context with similar-sized bodies in the Kuiper Belt.

As of the beginning of 2018, the KEM DKBO program is already targeting at least 24 DKBOs and Centaurs, including 10 cold classical KBOs, 5 dwarf planets, 3 Plutinos, 3 Centaurs, 2 scattered objects and 1 hot classical. These targets and various observing details about them are provided in Table 2 and Figure 6. These studies will provide key context with which to interpret both Pluto and MU69 results.

| Object | Date | Solar Phase Angle | Detectable Ring Fraction Relative to Chariklo* | Range, AU | Magnitude, Body | Magnitude, Body+Ring** | Class |
|---|---|---|---|---|---|---|---|
| 2011 JY31 | Oct-18 | 133 | 0.01 | 0.20 | 19.2 | 18.0 | Classical |
| 2011 JW31 | Oct-18 | 128 | 0.01 | 0.21 | 19.0 | 18.0 | Classical |
| 2011 HZ102 | Dec-18 | 126 | 0.02 | 0.21 | 19.1 | 18.0 | Classical |
| 2011 HK103 | Sep-18 | 125 | 0.02 | 0.34 | 19.1 | 18.0 | Classical |
| 1994 JR1 | Jul-16 | 129 | 0.04 | 0.61 | 20.0 | 18.0 | Plutino |
| 2011 HF103 | Dec-18 | 100 | 0.12 | 0.32 | 17.5 | 16.5 | Classical |
| 2011 JX31 | Jun-20 | 100 | 0.12 | 0.41 | 17.8 | 16.8 | Classical |
| 2014 PN70 | Mar-19 | 95 | 0.15 | 0.11 | 16.8 | 15.8 | Classical |
| Quaoar | Jan-19 | 84 | 0.26 | 11.91 | 18.9 | 17.9 | Classical |
| 2011 JA32 | Oct-18 | 107 | 0.28 | 0.50 | 20.0 | 18.0 | Classical |
| 2012 HE85 | Dec-17 | 80 | 0.30 | 0.32 | 17.2 | 16.2 | Classical |
| 2014 OS393 | Jan-19 | 78 | 0.33 | 0.09 | 15.6 | 14.6 | Classical |
| 2012 HZ84 | Dec-17 | 74 | 0.38 | 0.50 | 19.1 | 18.1 | Classical |
| 2002 KX14 | Jan-17 | 71 | 0.63 | 19.15 | 21.3 | 20.0 | Plutino |
| 2010 JJ124 | Jul-16 | 126 | 0.93 | 16.12 | 24.9 | 20.0 | Centaur |
| Huya | Jan-17 | 86 | 1.15 | 26.46 | 22.3 | 20.0 | Plutino |
| Pholus | Jan-17 | 119 | 2.10 | 15.75 | 25.1 | 20.0 | Centaur |
| Chiron | Jul-17 | 82 | 2.90 | 36.96 | 23.0 | 20.0 | Centaur |

**Table 2.** High-phase searches for forward-scattering rings or coma around KBOs and Centaurs. Key: *Minimum fraction of light seen from Earth that must be from the ring, relative to Chariklo (where the ring contributes ~1/3 of the light seen from Earth), in order for the ring to be detectable by NH, assuming scattering properties similar to Saturn's G ring. Smaller numbers imply greater sensitivity to ring material. **For a ring at the brightness sensitivity threshold.

The New Horizons team recently found that the spacecraft attitude stability is stable enough to permit LORRI observations as long as 30 sec, which improves the limiting sensitivity of LORRI by a factor of ~2-3 compared to previous limits (increasing the limiting magnitude for point sources from V~20.5 to V~21.5). We expect this new capability will enable observations of 4-10 additional DKBOs before KEM ends in 2021.

To demonstrate the utility of the DKBO program, the New Horizons LORRI instrument observed the Plutino 15810 Arawn (1994 JR$_1$) in late 2015 and early 2016 (Porter et al. 2016). These LORRI data were combined with ground-based and HST data to make the first determination of a KBO's solar phase curve of a KBO out to an angle of 58º (Figure 7). Hapke (2012) modeling of this disk-integrated solar phase curve was used to infer that Arawn has a rough surface with mean topographic slope angle of 37º±5º. LORRI rotational lightcurve data were also used to constrain its shape and determine for the first time that Arawn has a relatively fast rotational period of 5.47±0.33 hours (see Figure 7). Additionally, the improved astrometry obtained on Arawn demonstrates that this Plutino passes relatively close to Pluto every 2.4 million years, causing significant perturbations in Arawn's orbit (see also Marcos & Marcos 2012).

Similar results will be possible for many of the KBOs observed as part of the KEM DKBO program. LORRI photometry obtained at multiple epochs (phase angles) will be used to model the shapes of the small KBOs, as Cassini did for outer Saturn satellites (Denk & Mottola 2014). No shape determinations of KBOs this small had previously been obtained. As noted above, such measurements can be useful both to assess how typical MU69's shape is and to potentially constrain KBO origins—for example, if they are all nearly spherical, then pebble formation models or at least rubble pile structures might be preferred.

Furthermore, the DKBO solar phase curve measurements obtained during KEM will significantly constrain KBO surface micro-physical properties and will be important for interpreting KBO radiometric diameters and regolith directional scattering properties. Measurements at large solar phase angles are needed to characterize surface properties such as roughness, the single particle phase function, and the distribution of particle sizes because small particles are more forward scattering (see e.g., Helfenstein et al. 1988). Such data cannot be obtained presently for *any* KBOs owing to severe geometric limitations from Earth that only allow KBO solar phase measurements out to ~2°; New Horizons observations will extend KBO solar phase curve photometry to angles up to 130–140° in some cases, essentially two orders of magnitude higher than currently possible. This will also allow the first robust comparisons of KBO surface micro-physical properties to those of asteroids, Trojans, and icy satellites, comparison of Pluto's and Charon's phase curves to several other dwarf planets for the first time, and assessment of how typical MU69 is by comparing its solar phase curve to those of other small and large KBOs.

Further still, New Horizons will study at least seven objects passing closest to the spacecraft to explore the binary fraction of KBOs in two new ways: detecting satellites up to 10× closer to their primary and up to several times smaller than HST can (precisely how much depends on each KBOs distance at each observation epoch of it). Binarity is a powerful clue to understanding KBO accretion and is known to climb steeply at separations approaching the limit of HST's resolution (Kern & Elliot 2006). Many binary formation mechanisms prefer forming tight systems over wide ones (Funato et al. 2004; Schlichting & Sari 2008; Nesvorny et al. 2010; Porter and Grundy 2012), so it is conceivable that the binary fraction continues to rise to the limit of LORRI's resolution. Large Cold Classical KBOs (CCKBOs) are almost all binary (Noll et al. 2008). However, except for recent suggestions of binarity in MU69 itself (Buie et al. 2018), no binary has ever been detected among small CCKBOs, despite telescopic searches.

As a result, even the non-detection of binaries in the KEM sample will provide a new and statistically valuable measurement, limiting the binarity fraction to less than ~38% ($2\sigma$). LORRI's ability to also assess binary fraction, but at smaller separations than HST is also particularly powerful because these KBOs are relatively small targets compared to the majority of KBOs examined for binarity using HST. The binary fraction as a function of KBO size is a powerful tracer of collisional evolution (Petit & Mousis 2004; Nesvorny et al. 2010; Parker & Kavelaars 2012), with a collisionally evolved population predicted to have fewer binaries for smaller primaries if the size distribution of impactors is steeper than a critical power-law slope (Parker & Kavelaars 2012). Current observations suggest that the observed binarity rate drops with decreasing primary KBO size; KEM results will critically test this important formation mechanism. Individual phase functions of resolved components of detected binary CCKBOs in our sample will provide a new test for the common properties of their surfaces.

Finally, we note that because the KEM DKBO science benefits from groundbased and HST support observations at low- phase angles, particularly for phase function and rotational lightcurve investigations, the New Horizons team has organized an informal Earth-based and HST observation campaign to complement the New Horizons DKBO observation program.

**Searches for Chariklo-Like Rings Around KBOs and Centaurs.** The New Horizons KEM also provides a unique opportunity to observe KBOs and Centaurs at intermediate and high phase angles to determine the frequency of rings or other orbiting dust assemblages, as were recently and surprisingly found around the Centaurs Chariklo (Braga-Ribas et al. 2014) and Chiron (Ortiz et al. 2015) and the KBO Haumea (Ortiz et al. 2017). KEM accomplishes this by obtaining deep high phase imaging and photometric searches for rings of KBOs and Centaurs using LORRI.

To determine whether these types of orbiting assemblages are rare or routine, LORRI will examine at least 18 KBOs and Centaurs (Table 1; note that Haumea is a candidate for 2019-2020 observations, but Chariklo itself cannot be observed by LORRI owing to its small (≤10°) solar elongation angle during KEM).

The New Horizons team has examined the photometric detectability of rings or similar structures around all known Centaurs, KBOs, and Trojans by determining the potential brightness as seen by LORRI and compared this to the estimated brightness of the object itself at high phase without any rings. For this calculation, the rings were assumed to have photometric properties similar to those of Saturn's faint G ring. The results are given in Table 2. As shown there, LORRI can provide powerful constraints on ring systems around this group of various KBOs and Centaurs. In most cases, ring systems much fainter than Chariklo's can be detected if they are forward scattering. No other technique can practically rival the KBO/Centaur dust/ring survey scope KEM can achieve through 2021 when KEM concludes.

**Target of Opportunity DKBO Investigations.** Finally, we cite two kinds of target of opportunity (ToO) observations that may present themselves during KEM.

The first ToO is observing *KBO mutual events* that cannot be seen from Earth but can be seen from NH. Various key system parameters including the size and shape of both the primary and secondary bodies (and thus their bulk densities), their relative albedos, and even the presence of albedo features on their surfaces can be obtained from such events. We have searched for but have not yet found any such candidates, but we will annually re-visit this analysis using improved KBO and KBO-satellite orbits, also searching newly discovered systems. We plan to schedule LORRI observations as opportunities arise.

The second ToO regards *KBO stellar occultations or appulses*. Once Gaia data releases improve the orbital and ephemeris fidelity of most known KBOs in 2018–2019, we will conduct a thorough search for both visible (LORRI) and UV (Alice) stellar appulse candidates observable from NH in order to find opportunities to probe them for coma, rings, and other opacity sources. Identified opportunities with good signal-to-noise ratio (SNR) stars would provide a statistical context in which to place our planned coma searches around MU69 (see §5), and will be scheduled for observation if found.

## 5. The Flyby Reconnaissance of 2014 MU69

The close flyby of the CCKBO MU69 is the centerpiece of the New Horizons extended mission. This flyby, which will culminate in a close approach near 5:33 UT on 1 January 2019, will provide the first and only opportunity in the foreseeable future for close-up observations of a small, primitive Kuiper Belt object. Those observations will by extension provide knowledge that can be extrapolated to better inform out understanding of other small KBOs. Here we summarize the planned exploration of MU69.

**Discovery and Selection of MU69 as a Flyby Target.** Because no previously known KBOs beyond Pluto were accessible to New Horizons, a dedicated search for flyby targets was necessary. A ground-based search, predominantly using the Subaru and Magellan telescopes, discovered about 60 new KBOs in the vicinity of New Horizons' trajectory, but none was close enough to be reachable by the spacecraft, and a trajectory deflection to the closest one would have required about twice as much propellant as was available. The search depth and discovery rate was limited by the extremely high background star density due to the low galactic latitude of the search area. Therefore in 2014 we obtained 160 orbits of Guest Observer time, supplemented by 40 orbits of Director's Discretionary time, with HST for a deeper search using the Wide Field Camera 3 (WFC3). Due to the background star confusion, the high angular resolution of HST provides an improved limiting magnitude, V~27.5, compared to the V~26 limit for the Earth based surveys. The HST search yielded two targetable KBOs, 2014 MU69 and 2014 PN70. MU69 was chosen as the flyby target because the propellant required for targeting it was roughly half that required for targeting 2014 PN70, leaving ample reserves for other science activities.

**Science Objectives**. MU69 is a member of the Cold Classical Kuiper Belt (CCKB), a structure of bodies in long-term, stable, nearly circular, low-inclination orbits that likely represent the only surviving remnants of the proto-planetary disk that have not been significantly thermally altered since the formation of the Solar System (Dawson & Murray-Clay 2012).

The geology of MU69 is expected to reveal much about its origin and evolution, and how this primitive mid-sized object's landforms and regolith compare to those of comets, asteroids, and icy satellites. Its overall shape, including whether it is a contact or closely orbiting binary, as suggested by stellar occultation results (Buie et al. 2018SZ), and the presence or absence of satellites, will be a key constraint on planetary accretion models. New Horizons will make imaging measurements to search for geologic and regolith units, as well differing surface unit ages via the determination of the spatial variation in crater populations across the surface. New Horizons will also search for layering on exposed scarps, crater walls, and other topographic features as clues to the accretion of cold classical KBOs, to search for evidence of interior or surface fragmentation, and to search for evidence of surface and interior evolution, including possible volatile loss. Crater populations will also be used to constrain the collisional history of the Kuiper Belt. Fracture patterns and topography will be used to constrain internal strength and history. Slope measurements will be used to constrain the local stability and cohesion (e.g., sintering) of granular materials on MU69's surface. New Horizons will also search for rings, analogous to those found recently around Chariklo and Haumea (Braga-Ribas et al. 2014; Ortiz et al. 2017). If detected, New Horizons would provide the first close-up observations of a small body ring system, potentially providing key information on the formation and evolution of these puzzling structures.

Studies of MU69's color and composition will illuminate the thermal and compositional environment and the nature of the material from which it was assembled, radial mixing in the nebula (e.g., Dalle Ore et al. 2013), and the evolution of MU69's surface. Because of its long-term cold storage, some fraction of MU69's constituents may have survived little altered from icy dust particles and trapped gases from the proto-solar nebula. Another fraction may consist of refractory materials that were thermally processed close to the proto-Sun and then radially mixed outward (e.g., Wooden 2008). Between these extremes may be solids resulting from varying degrees of thermal processing, potentially including sublimation and condensation of some volatiles but not others.

New Horizons observations will search for subunits across MU69 by searching for differences in their visible colors and by detecting infrared absorption features in the 1.25-2.50 µm band that the Ralph LEISA IR mapping spectrometer covers. In addition to searching for surface color and composition variegation, New Horizons will also search for regolith vertical compositional heterogeneity using craters as windows to the interior. These various studies may also shed light on the puzzling color diversity of KBOs in general, e.g., if differing color units are found on MU69. Compositional and color data on MU69 could establish links between CCKBOs and the emerging compositional classes of comets (e.g., Mumma & Charnley 2011; A'Hearn et al. 2012). Accretional history may be particularly evident if MU69 is a collisional fragment because the fragmentation may have exposed the internal compositional and physical structure of its parent body.

Because of its small size and stable orbit, it is likely that active volatiles that might have once been present on or near MU69's surface have long ago escaped to space, leaving only inert ices, tholins, minerals, or lag deposits, which would not produce a coma at MU69's large distance and resulting low surface temperature. However, conventional wisdom may not be correct, as first flybys of other targets (and ground-based studies of main belt comets, e.g., Jewitt et al. 2012) have sometimes revealed. Therefore, the New Horizons MU69 flyby will be used to search for evidence of atmosphere/coma and dust around MU69, both directly via ultraviolet emission and absorption searches and high phase angle visible wavelength imaging, and also indirectly by searching for plasma interactions between MU69 and the solar wind.

Table 3 summarizes the key science and measurement objectives for the MU69 flyby, which are designed to address the science objectives outlined above. Table 3 also details the expected data quality. The flyby trajectory selection and a summary of planned observations are provided below.

**Table 3: New Horizons MU69 Primary Science and Measurement Objectives**
(Quantitative measurements refer to the prime trajectory and flyby sequence)

| Science Objective | Measurement Objective | Best Achieved at MU69 in Prime Sequence (Green: better than Pluto; Blue: comparable to Pluto; Red: less good than Pluto) |
|---|---|---|
| **Group 1** | | |
| Characterize the global geology and morphology | Panchromatic full-disk close approach imaging | Full disk 0.13 km/pixel, partial coverage at 0.035 km/pix |
| | Panchromatic rotational coverage imaging | 1.5 km/pix or better, all longitudes* |
| | Topography, digital elevation models (DEM) | 0.14 km/pixel |
| Map surface composition | Close approach full-disk IR maps | 1.8 km/pix |
| | IR compositional spectroscopic rotational coverage | 19 km/pix or better, all longitudes* |
| | Color full-disk close approach imaging | 0.33 km/pix |
| | Color rotational coverage imaging | 6 km/pix or better, all longitudes* |
| | UV reflectance spectroscopy | Disk integrated |
| Search for satellites and rings | Deep high and low phase imaging for satellite and ring searches | Nested observations starting with full Hill sphere; most sensitive satellite detection threshold ~0.2 km diam to 5000 km distance (~5% of Hill sphere), ring detection threshold I/F ~5e-7 |
| **Group 2** | | |
| Characterize composition and magnitude of any volatile or dust escape | UV stellar occultation coma search | Alice stellar appulse |
| | UV solar occultation coma search | Alice solar appulse (0.5 hour integration) |
| | UV coma airglow search | Several Alice scans of potential coma, high and low phase |
| | Heliospheric Ly-alpha coma absorption search | Alice scans near C/A |
| | High-phase imaging dust coma search | Multiple high-phase coma dust searches |
| | Volatile escape detection via plasma interaction | Near-continuous SWAP and PEPSSI measurements near MU69, including plasma rolls |
| Surface properties of MU69 | 4 cm day and night brightness temperature | Both day and night, hemispheric unresolved |
| | Near-IR spectroscopic temperature measurements via band shifts | 1.8 km/pix |
| | Range of phase angles for MU69 to determine photometric properties | Global color and pan imaging at phase angles 1.5 - 169 degrees |
| | Photometric properties of any satellites and rings | Global color and pan imaging at phase angles 4.5 - 169 degrees |
| Crater size/frequency distributions | High-resolution imaging | Full disk 0.13 km/pix, partial coverage at 0.035 km/pix |
| Characterize any satellites and rings | Sizes, shapes, rotation periods of any satellites | 0.6 km/pix to 1,000 km distance, dozens of visits for lightcurves |
| | Geology of any satellites | 0.6 km/pix to 1,000 km distance |
| | Color of any satellites and rings | 2.5 km/pix to 10,000 km distance |
| | Surface composition of any satellites and rings | 6 km/pix to 1,200 km distance |

* For typical rotation period of 8 hours

**Flyby Trajectories**. The choice of flyby trajectory at MU69 was a fundamental constraint on the science to be obtained during the flyby. The New Horizons team selected two close approach points: a prime, to be flown in the absence of navigation difficulties or hazards close to MU69; and an alternate which can be employed if navigation difficulties or hazards warrant a more conservative, distant point. The choice between the prime (closer) and alternate (more distant) closest approach points will be made in late 2018 on final approach to MU69.

The speed and direction of approach to MU69 was fixed by the respective orbits of the spacecraft and the KBO. The flyby relative speed will be 14.4 km s$^{-1}$; the asymptotic approach direction is approximately in the ecliptic plane at an angle of 11.7 degrees from the direction of the Sun. The position of the spacecraft relative to MU69 at closest approach, in a plane perpendicular to the approach direction (called the B-plane), was a free parameter that was chosen to optimize the flyby science. This position can be specified by the position angle $\varphi$ of the vector from MU69 to the spacecraft at closest approach (defined such that $\varphi = 0$ when the angle between this vector and the direction of the Sun, as seen from MU69, is minimized), and the magnitude of that vector (i.e., the closest approach distance).

The close approach distance was selected as a compromise to optimize the overall flyby science between various competing in situ and remote sensing objectives. A closer flyby generally improves spatial resolution, and the chance of detecting the influence of MU69 on its heliospheric particle and dust environment by in situ methods. However, a more distant flyby provides more time for observations at each particular viewing geometry, because the geometry changes more slowly during the flyby, which is important when optimizing lighting for remote sensing or for stereo imaging. A more distant flyby also reduces the likelihood that images will be smeared near closest approach or will miss the target due to pointing uncertainties, and is likely to reduce the risk to the spacecraft from collisions with potential debris surrounding the target.

Similarly, there are competing constraints on the choice of $\varphi$. Because nothing is known in advance of the flyby concerning the position angle of MU69's spin axis, its surface features, or the existence or orbital positions of any satellites, the only factors that can be usefully controlled in designing the flyby are the viewing direction relative to the Sun, which controls the illumination of the visible surface and the apparent position of MU69 relative to the Sun and background stars. The end-member orientation choices are $\varphi = 0°$, where the spacecraft flies through the line connecting MU69 to the Sun prior to encounter, and $\varphi = 180°$, when the spacecraft flies through the shadow of MU69 after the encounter. The phase angle (i.e., the spacecraft-target-Sun angle) at any time relative to closest approach is larger for larger $\varphi$. The angle $\varphi$ can also be adjusted to change MU69's apparent position relative to stars to optimize optical navigation on approach, and to adjust the target's position relative to the Sun and other stars to probe the possible presence of gas or dust near MU69 using solar and stellar occultation techniques.

Another variable to be specified, in tandem with the trajectory, is the uncertainty in MU69's position to be covered by the encounter observations. The estimated 1σ uncertainty in MU69's position prior to encounter is ~30 km perpendicular to the approach direction, and ~600 km along-track, i.e. parallel to that direction. Close approach observations are designed to cover these uncertainties.

The prime MU69 flyby trajectory is designed for a 3,500 km closest approach distance, in ecliptic coordinates to the north of MU69, and with $\varphi$ = 82°. This closest approach distance was primarily selected as a compromise between maximizing spatial resolution, and minimizing both smear and the chance of missing the target (due to navigation uncertainties) in the highest-resolution image sequences. The choice of $\varphi$ balanced the desires for (i) relatively high phase angles during the closest images, (ii) optimized illumination of topographic features, (iii) relatively low phase angles to maximize signal-to-noise ratios for spectroscopy on approach, while (iv) also minimizing the interference of bright background stars with optical navigation images on approach. The prime close approach sequence adopted an along-track coverage for the highest priority observations of ±3900 km (6.4σ); this generous coverage provides substantial robustness against potential underestimation of along-track uncertainties.

The alternate MU69 flyby trajectory is designed to a 10,000 km closest approach distance and $\varphi$ = 92°, also with close approach to the north of MU69. The closest approach distance, about three times as far from MU69 as the prime trajectory was selected to be as close as possible while protecting against estimated hazards and navigation uncertainties that could negate flying the prime trajectory. The choice of $\varphi$ on the alternate trajectory was made for the same reasons as in the prime trajectory. An along-track uncertainty coverage of ±7,200 km was to allow for an increased uncertainty margin in MU69's position.

For both trajectories, we chose the nominal close approach time to be 05:33 UT on 1 January 2019. The flyby date was selected to minimize propellant consumption. The flyby time on that date is about an hour earlier than the uncontrolled (minimal propellant usage) arrival time, to allow the spacecraft to receive uplink signals from both the NASA Canberra and Goldstone Deep Space Network (DSN) stations to attempt a bistatic radar measurement of MU69 shortly after closest approach.

**Planned Flyby Observations.** Unresolved MU69 approach observations will begin in late August 2018, with a first attempted detection of MU69 using the LORRI instrument. Successful detection of MU69 is guaranteed by late September 2018, ~100 days before closest approach and at an expected visual magnitude near 19.5. These LORRI images will be used primarily to provide optical navigation data, but also to search for both satellites and rings. Regarding rings and satellite searches, a more intensive campaign between 32 and 19 days before encounter will return deep (limiting visual magnitude ~21.5) images of the environment around MU69 to search for potentially hazardous dust structures, or satellites capable of generating hazardous material. These data will be used to inform the decision on whether to divert to the hazard-avoiding alternate trajectory and sequence, which can be selected as late as 10 days before closest approach. Detected satellites will be tracked to determine their orbits, to obtain more detailed imaging, and to determine the offset of the MU69 primary from the system barycenter to better target close approach observations of the primary itself.

Throughout the approach to MU69, SWAP, PEPSSI, and SDC will observe the heliospheric plasma (solar wind, pick-up ions, and energetic particles) and the dust environment near MU69.

MU69 will subtend 2 LORRI pixels about 2.5 days before closest approach. New Horizons will obtain resolved images from this point inward to study the rotationally resolved photometric, color, and geological properties of MU69. Starting about 1 day before closest approach, Ralph color images, Ralph near-IR composition spectroscopy, and Alice ultraviolet spectroscopy will be interleaved with the LORRI images to provide multi-wavelength rotational coverage. Deep searches for satellites and rings will also continue, ultimately being capable of detecting satellites as small as 100 meters diameter for a characteristic visible albedo of 0.1.

About 3 hours before closest approach, New Horizons will obtain images and spectra blanketing a region of about 1000 km radius around MU69 to characterize the shapes and compositions of any satellites or rings in this region. A re-targetable observation, with pointing to be finalized as late as 3 days before closest approach, will allow high resolution satellite imaging if there is a satellite whose position is sufficiently well known to be targeted. Numerous Alice UV spectrograph integrations on approach will search for airglow emissions from any gaseous coma surrounding MU69.

About 1.25 hours before closest approach, the expected uncertainty in MU69's apparent position will become large enough that scans are required to cover the possible range of positions (see Figure 8). For the prime trajectory, color imaging and panchromatic imaging with MVIC, and near-IR spectroscopy with LEISA will be obtained through closest approach, with best resolution being 300 m/pixel, 140 m/pixel, and 1.8 km/pixel, respectively. The LORRI framing camera, riding along with these scans, will obtain images with best resolution 35 m/pixel on the prime trajectory, though most LORRI images will have greater smear and lower SNR than the MVIC images. Images from multiple viewing directions, with phase angles up to 165 degrees, will be used to characterize surface photometric properties, and to obtain digital elevation maps using stereo techniques. New Horizons will also make disk-integrated observations of day- and night-side microwave thermal emission using REX, and UV surface reflectance using Alice. Working with NASA's Deep Space Network (DSN), REX will also attempt an uplink bistatic radar detection reflected from MU69's surface. Plasma and dust observations with SWAP, PEPSSI, and SDC will be conducted during the close approach to search for dust and plasma interactions.

On departure, the Alice ultraviolet spectrograph will observe the Sun to search for any absorption due to possible coma gases traversed by the sunlight on its way to the spacecraft, and will similarly probe the possible coma via a stellar occultation. Subsequent departure observations will conduct deep searches for forward-scattering rings around MU69, and will attempt high-phase imaging of MU69 itself, though its surface will rapidly become undetectable in the Sun's glare as the spacecraft recedes, due to the high departure phase angle of 169 degrees.

The alternate flyby sequence (10,000 km closest approach) will obtain similar observations, but best resolutions will be about a factor of 2 lower than for the prime trajectory.

Approximately 50 gigabits of data will be collected during the MU69 flyby, regardless of which trajectory is ultimately used. Owing to low data rates from the Kuiper Belt and the shared nature of the DSN, these data will require approximately 20 months to be downlinked to Earth.

## 6. Prospects for Future New Horizons Extended Missions

Twelve years after launch, and almost halfway through its first extended mission, New Horizons and all 7 instruments aboard it remain healthy; all of the prime and backup systems are fully functional.

The telecommunications system aboard New Horizons is capable of communicating with the current DSN out to distances in excess of 200 AU, a distance that won't be reached until ~2070.

Power is the ultimate limiting driver in mission lifetime if the spacecraft is simply spin stabilized for (SWAP and PEPSSI) energetic particles, (Alice) UV neutral gas, and SDC (and possible REX) dust impact rate studies of the heliosphere. Had the RTG aboard New Horizons received a complete load of plutonium, another decade of life would have been expected from the additional 30 watts at beginning of life, but the Los Alamos fuel production line shutdowns of 2004-2005 resulted in New Horizons launching with only about 90% of a full plutonium supply. Given the 3.3 watt/year power production decline owing to the plutonium half-life, the spacecraft propellant and power likely will limit operations to ~2035 or perhaps a few years later. This will limit the heliospheric range achievable to 90-110 AU, depending on how far engineers can extend the lowest power mode of the spacecraft and its instrument payload. Voyagers 1 and 2 crossed the termination shocks at ~94 and 84 AU respectively. Therefore, the New Horizons power budget likely enables particle observations, extending to or even past the termination shock (Richardson and Stone, 2009).

Nevertheless, throughout the 2020s, there is significant potential for valuable planetary science (and even astrophysics), with New Horizons. Models indicate some 7-10 kg of usable hydrazine (not including the estimated trapped, ~2 kg) will remain at the end of the first extended mission to facilitate unique observations.

Although this remaining propellant is not sufficient to target any known or foreseeably known KBO close flyby, there is the possibility to use the LORRI camera aboard New Horizons to search for comet nuclei near its path and to target such a flyby if one is located. The flyby of a thermally unprocessed comet nuclei one to several km in diameter would be an important experiment not achievable by any other means.

Additionally, the useable propellant remaining aboard after the first extended mission is capable of supporting hundreds of pointed observations of KBOs, dwarf planets, Centaurs, other planetary targets at high phase (e.g., the Saturn Phoebe ring). Candidate observations types include:

- High phase monitoring of Triton's atmosphere and surface changes.
- High phase monitoring of Enceladus' plumes.
- High phase observations of Haumea's rings.
- A complete survey of all Centaurs for evidence of ring systems detectable at high phase.
- Observations of additional small KBOs near the flight path to study shapes, lightcurves satellite populations, and surface regolith properties by moderate and high phase angle photometry.
- Sampling of very small KBOs using LORRI to probe their size-frequency luminosity distribution of KBO, as a function of distance.
- UV monitoring of Jovian and Saturnian auroral luminosity.
- KBO UV and visible stellar occultations not observable from Earth.
- Dust density measurements as a function of heliocentric distance using SDC and possibly REX.

Additionally, potential astrophysical observation types include:

- Long-term UV monitoring of stellar and other sources not feasible with HST.
- Photometric studies of transient novae, kilo-novae, supernovae, and other sources (e.g., LIGO alert sources of black hole and neutron star mergers) not visible in the terrestrial daytime sky when NH and the Earth are on opposite sides of the Sun each year.

➢ UV and visible extragalactic background light (EGBL) photometry made feasible by being beyond all significant zodiacal light sources.

➢ REX measurements of both micro-lensing and gravitational waves using VLBI techniques to observe Quasi-Stellar Objects (QSOs),

These exciting prospects suggest the following notional future for New Horizons after its current KEM Extended Mission 1 (EM1) completes in 2021 some 50 AU from the Sun:

➢ EM2 (2022-2024, 50-60 AU heliocentric distance): Planetary astronomy of the Kuiper Belt, Cecntaurs, and other targets, possible primordial comet nucleus flyby, and continued in situ heliospheric surveys.

➢ EM3 (2025-2026, 60-70 AU heliocentric distance): Astrophysical observatory mission taking advantage of the 50+ AU range, a 6-month per year transient monitoring of terrestrial daytime objects of high priority, and continued in situ heliospheric surveys. The reason these observations are best carried out here is so they occur after the planned final planetary science in EM2.

➢ EM4 (2027-2035+, 70-90+ AU heliocentric distance): Heliospheric charged particle, dust, and neutral gas mission to probe the termination shock and possibly the ISM (depending on the ISM boundary distance during these years as it fluctuates with solar cycle).

## 7. Summary

NASA's New Horizons has completed its prime mission to make the first exploration of Pluto and its system of moons and has been approved for its first extended mission, which has the objective of extensively studying the Kuiper Belt environment, observing numerous Kuiper Belt Objects and Centaurs in unique ways, as well as making the first close flyby of the KBO 486958 2014 MU$_{69}$. We have summarized the objectives and plans for this mission extension, and briefly looked forward to potential objectives for subsequent extended missions by New Horizons.

## Acknowledgements

This work was supported by NASA New Horizons funding. We thank two anonymous referees for close readings of this review and their many useful comments.

# Figures

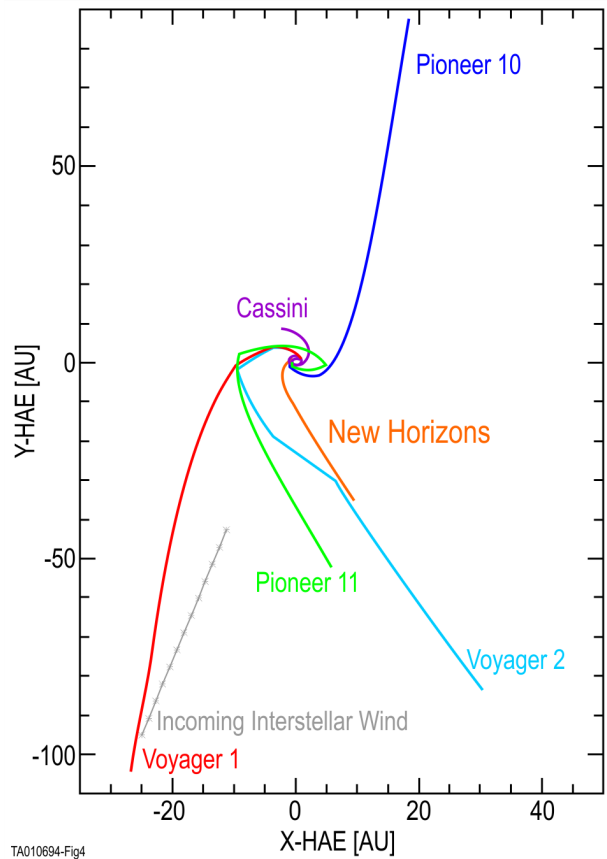

**Figure 1: Trajectories of New Horizons and previous outer heliospheric missions** (Elliott et al. 2016). Here HAE is the Heliospheric Aries Ecliptic Coordinates, In this system the Z-axis is normal to and northward from the ecliptic plane; the X-axis extends toward the first point of Aries (the Vernal Equinox, i.e. to the Sun from Earth in the first day of Spring). The Y-axis completes the right-handed coordinate system.

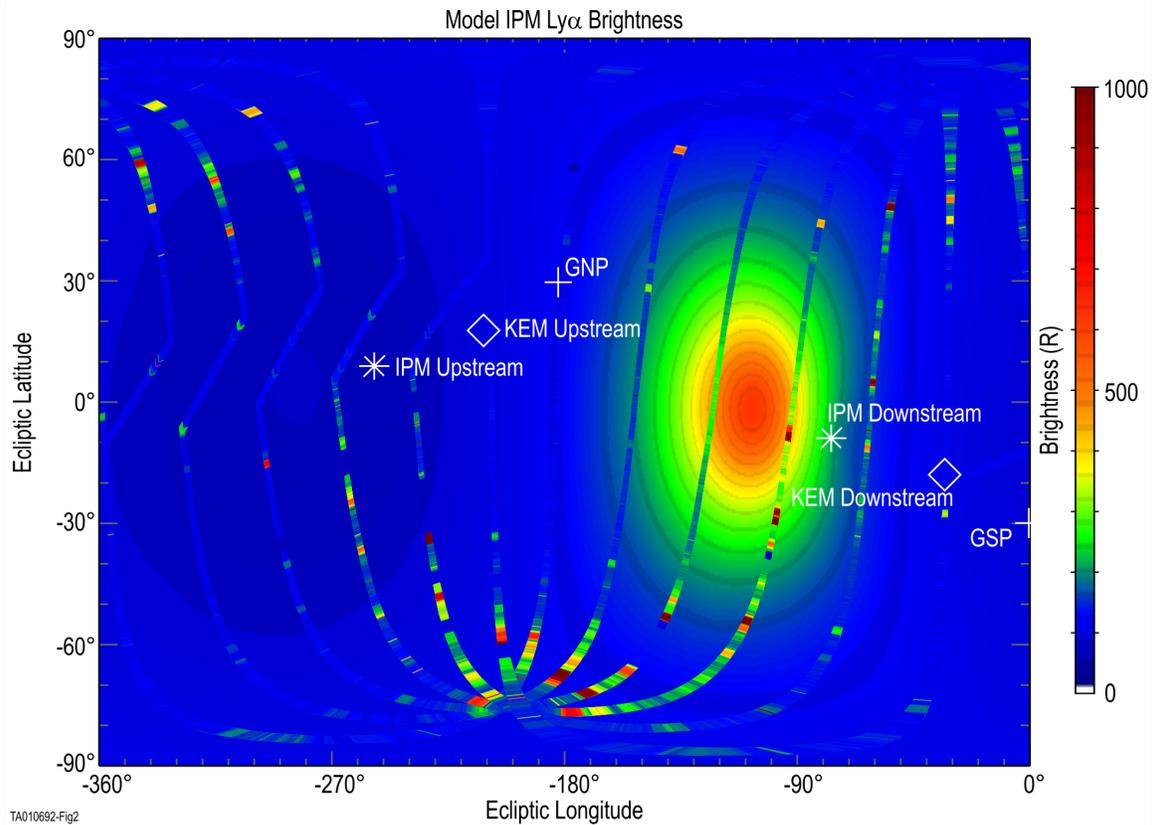

**Figure 2. Model all-sky brightness of interplanetary Lyα emission**. Brightness here is estimated at Pluto's location at the epoch of NH flyby, displayed in ecliptic coordinates and overlaid with New Horizons Alice UV data from the Pluto flyby, which is similar to that planned for KEM depicted as the great circle "stripes" every 30 degrees. The model brightness peaks in the direction of the Sun; the sunward hemisphere average interplanetary Lyα brightness of 195 R, or about twice the anti-sunward hemisphere average interplanetary Lyα brightness of 95 R. The location of the upstream and downstream Interplanetary Medium (IPM) directions and the Galactic North (GNP) and South (GSP) poles are labeled from (Gladstone et al. 2018).

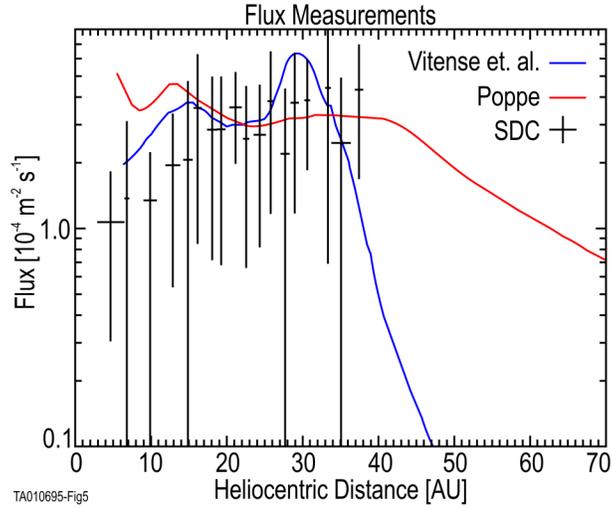

**Figure 3. Comparison of SDC observed dust fluxes to recent models**. Notice the models dramatically diverge beyond ~35 AU: blue (Vitense et al. 2014) and red (Poppe 2016). KEM data returned by SDC will discriminate between these divergent models (Piquette et al. 2017).

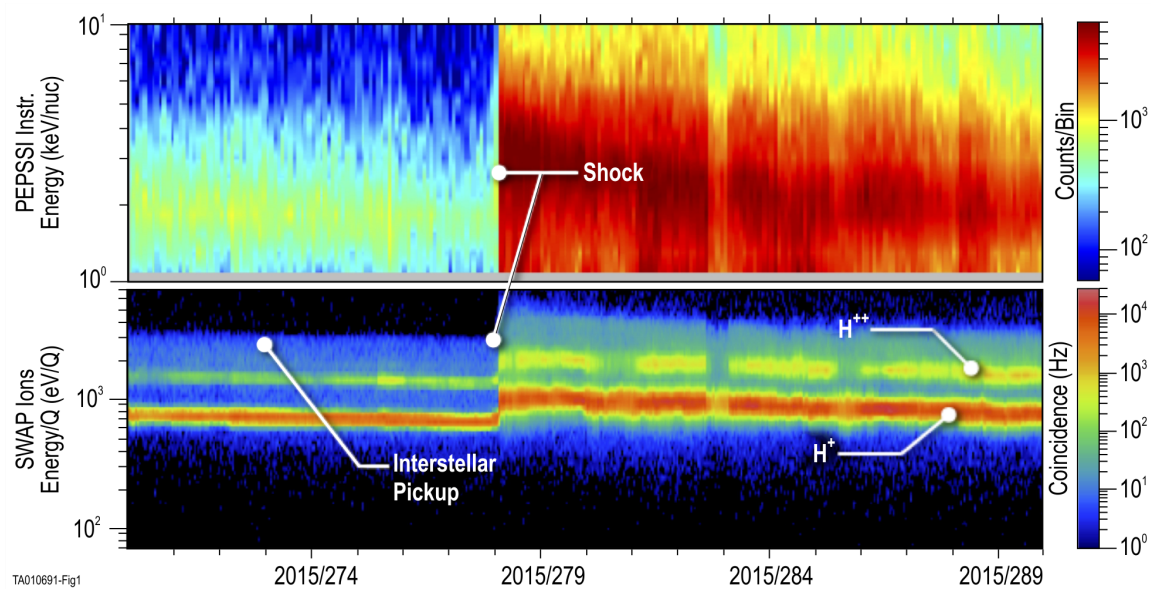

**Figure 4: Sample PEPSSI and SWAP plasma outer heliospheric observations.** PEPSSI (upper panel) and SWAP (lower panel) easily see the time-varying plasma properties of the outer heliosphere from keV energies by SWAP detecting solar wind $H^+$, $He^{++}$, $H^+$ Pick Up Ions (PUIs), to MeV energies with PEPSSI observing $He^+$, PUIs, and ions of H, He, CNO suprathermals. Intensities are modulated in part by solar activity (e.g., by the passage of a strong shock produced by a CME).

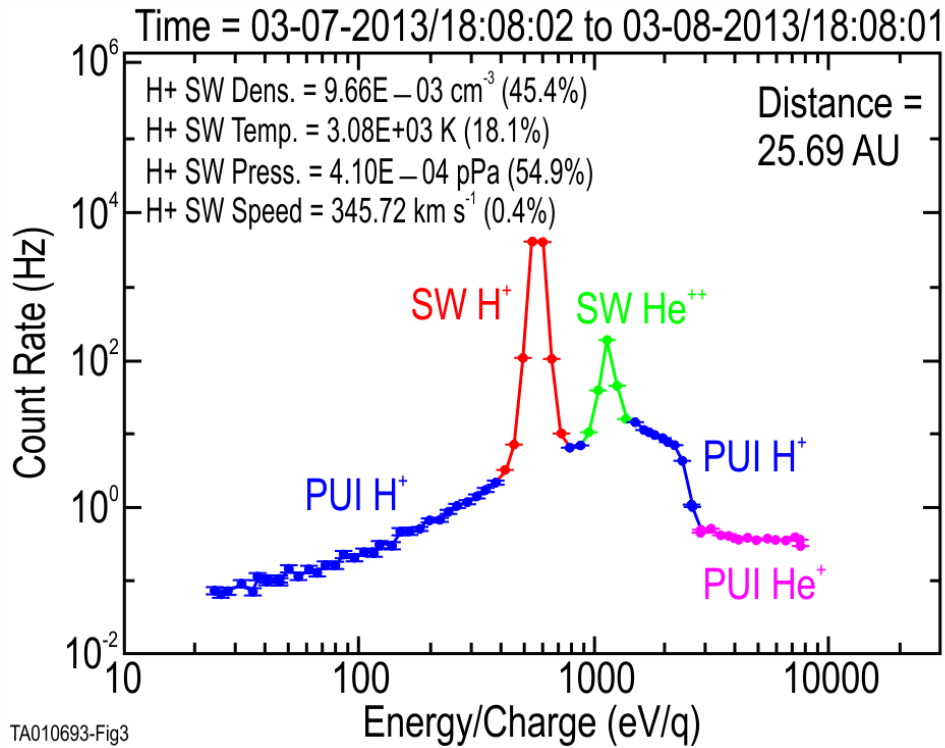

**Figure 5: SWAP count rate-energy per charge distribution at ~25.7 AU** (McComas et al. 2017). The plot is annotated with the primary source of the counts: solar wind (SW) or interstellar PUIs. The distribution is a 1-day time average; error bars are show. The average solar wind conditions are shown in the upper left corner. Percentages in brackets on the solar wind parameters are normalized root mean square (RMS) variations of hourly values.

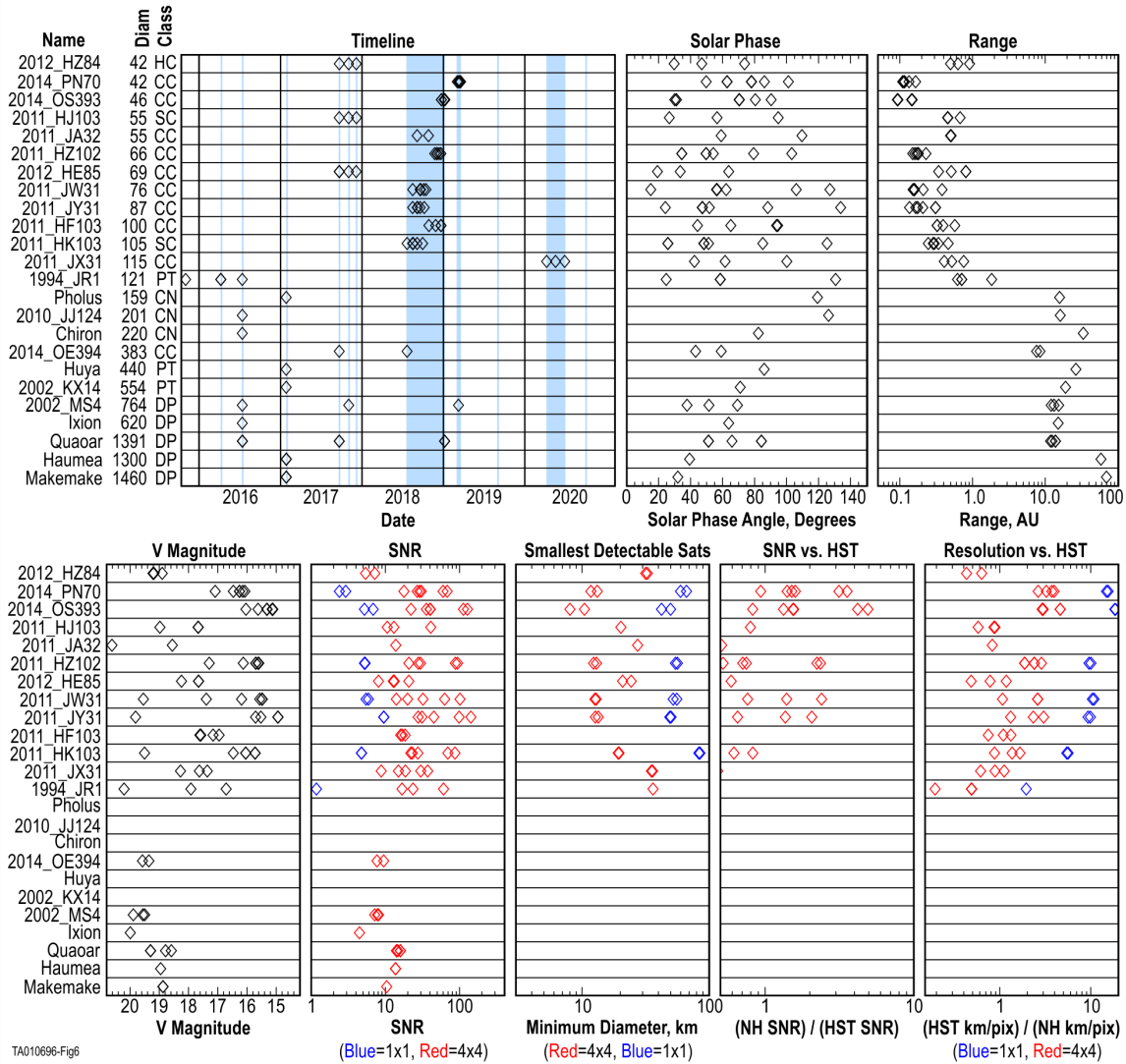

**Figure 6. Summary of the distant KBO and Centaur observation plans as of 1 Jan 2018.** In the timeline (upper left), blue vertical bars indicate operational periods when observations are possible. Object diameters, in km, assume albedos of 0.1 for objects where the true albedo is not known. Object classes are as follows: CC=Cold Classical; HC=Hot Classical; CN=Centaur; PT=Plutino; SC=Scattered; DP=Dwarf Planet. Diamonds show geometry and expected observation SNRs; spatial resolution and SNR are compared to HST (for a single HST orbit assuming WFC3/F350LP). 4×4 binned images (red) have higher SNR but lower spatial resolution, and allow searches for smaller satellites than are otherwise possible; unbinned 1×1 images (blue) search for closer-in satellites and binaries than are otherwise possible. Satellite search limits here assume an albedo of 0.1 and a 3σ detection threshold. Note that Pholus, 2010_JJ124, Chiron, Huya, and 2002_KX14 do not appear in the lower panels, because they are only detectable if they have extensive forward-scattering ring material.

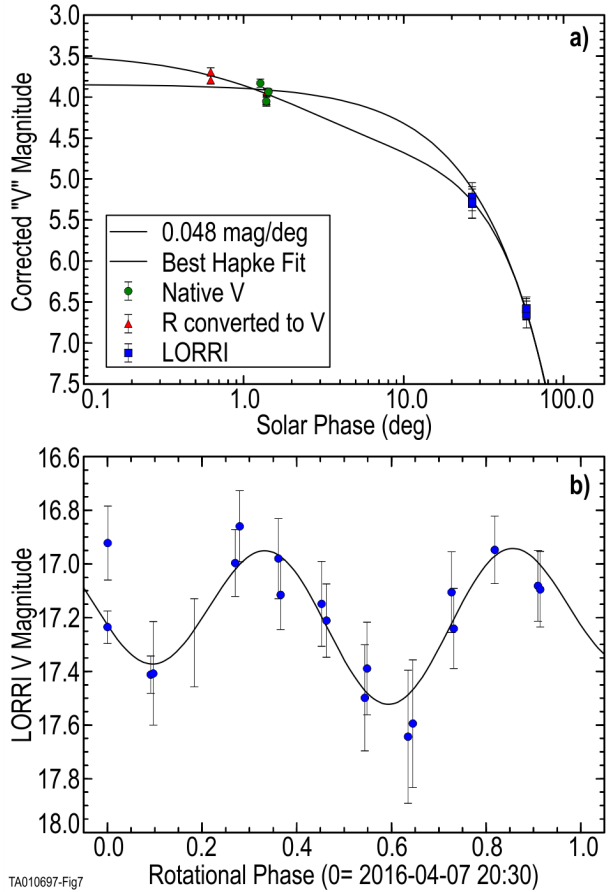

**Figure 7. LORRI photometry of the KBO Plutino (15810) 1994 JR$_1$ (Arwan).** Top panel: The best-fit 1994 JR$_1$ phase curve and points from LORRI, Green et al. (1997), Benecchi et al. (2011), and D. Tholen (priv. comm.). R-band photometry was converted to V-band with V–R=0.76. Bottom panel: 2016 April LORRI Arwan data phase-folded over a period of 5.47 hour, with median V=16.9 and peak-to-trough amplitude of 0.8 mag (adapted from Porter et al. 2016).

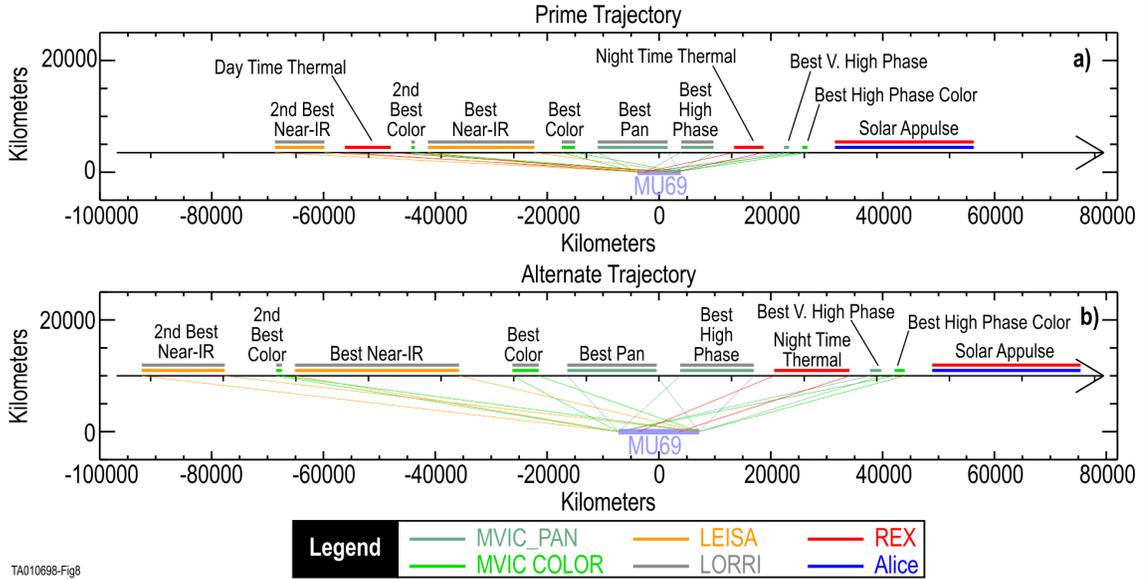

**Figure 8. Simplified diagram of the prime (upper panel) and alternate (lower panel) MU69 close approach sequences.** Times are approximate, and additional Alice, SWAP, PEPSSI and SDC observations are omitted for clarity. Tick marks along the trajectory show 15-minute time intervals, and the diagonal lines show the view direction from the spacecraft to MU69. The lower bar shows the maximum uncertainty in MU69's position that will be covered by the observations. Ecliptic north is up, and the Sun illuminates MU69 from the left, 11 degrees out of the plane of the diagram.